\documentclass[fleqn,usenatbib]{mnras}

\usepackage{newtxtext,newtxmath}

\usepackage[T1]{fontenc}

\DeclareRobustCommand{\VAN}[3]{#2}
\let\VANthebibliography\thebibliography
\def\thebibliography{\DeclareRobustCommand{\VAN}[3]{##3}\VANthebibliography}

\providecommand{\e}[1]{\ensuremath{\times 10^{#1}}}

\usepackage{graphicx}	
\usepackage{amsmath}	
\usepackage{multirow}
\usepackage{mhchem}
\usepackage{ulem}

\title[Low-temp. hydrogenation of aromatic archetypes]{Hydrogenation of small aromatic heterocycles at low temperatures}

\author[Miksch et al.]{April M. Miksch,$^{1}$ 
Annalena Riffelt,$^{1}$
Ricardo Oliveira,$^{2}$
Johannes K\"astner,$^{1}$ and
Germ\'an Molpeceres,$^{1}$\thanks{E-mail: molpeceres@theochem.uni-stuttgart.de}\\
$^{1}$Institute for Theoretical Chemistry, University of Stuttgart, Stuttgart, Germany\\
$^{2}$Chemistry Institute, Federal University of Rio de Janeiro, Rio de Janeiro, Brazil\\
}

\date{Accepted XXX. Received YYY; in original form ZZZ}

\pubyear{2021}

\begin{document}
\label{firstpage}
\pagerange{\pageref{firstpage}--\pageref{lastpage}}
\maketitle

\begin{abstract}
The recent wave of detections of interstellar aromatic molecules has sparked interest in the chemical behavior of aromatic molecules under astrophysical conditions. In most cases, these detections have been made through chemically related molecules, called proxies, that implicitly indicate the presence of a parent molecule. In this study, we present the results of the theoretical evaluation of the hydrogenation reactions of different aromatic molecules (benzene, pyridine, pyrrole, furan, thiophene, silabenzene, and phosphorine). The viability of these reactions allows us to evaluate the resilience of these molecules to the most important reducing agent in the interstellar medium, the hydrogen atom (H). All significant reactions are exothermic and most of them present activation barriers, which are, in several cases, overcome by quantum tunneling. Instanton reaction rate constants are provided between 50~K and 500~K. For the most efficiently formed radicals, a second hydrogenation step has been studied. We propose that hydrogenated derivatives of furan, pyrrole, and specially 2,3-dihydropyrrole, 2,5-dihydropyrrole, 2,3-dihydrofuran, and 2,5-dihydrofuran are promising candidates for future interstellar detections.

\end{abstract}




\section{Introduction}

Aromatic chemistry happening in astrophysical environments is puzzling for one particular reason: While the beginning of organic chemistry as a discipline can be arguably attributed to the work of August Kekul\'e on the structure of benzene (the building block of aromatic molecules), the study of aromatic chemistry in interstellar environments is much less common. This fact is indeed surprising when considering that aromatic molecules are more stable than their aliphatic counterparts due to the electron delocalization in the structure, and, thus, the prevalence of these molecules should be high in astronomical environments. The nature of the electronic structure of benzene still receives significant attention in the literature \citep{Liu2020, Eriksen2020}. 

The detection of circumstellar benzene \citep{Cernicharo2001, Malek2011}, as well as the reliable detection of benzonitrile (\ce{C6H5CN}) and cyanonapthalene \citep{McGuire2018, McGuire2021} in the TMC-1 molecular cloud has initiated a new wave of detections that are related to aromatic chemistry. The detection of benzonitrile in other sources \citep{Burkhardt2021}, as an example, proves that TMC-1 is not a particular case. The detection of other non-armoatic cyclic species that can share part of their chemical formation routes \citep{McCarthy2020, Lee2021} with aromatic compounds as well as the detection of proxies which are thought to be involved in the synthesis of aromatic molecules \citep{Agundez2021, Marcelino2021} also contribute to the growth in knowledge concerning aromatic chemistry. In addition to the detection of these molecules, it is important to mention the detection of \ce{C60+} in diffuse media \citep{Campbell2015}. The importance of these detections lies in the evident interest of knowing that these molecules populate dense molecular clouds and the possible synthetic pathways that account for the formation of these molecules in a bottom-up approach from smaller unsaturated precursors. We highlight the recent study of the formation of indene in this context \citep{Doddipatla2021}.  The bottom-up approach contrasts with the top-bottom one, which considers aromatic molecules as products of the energetic processing of large polycyclic aromatic hydrocarbons or soot-like structures \citep{Tielens2008, Merino2014}. This scenario presents an alternative explanation for the formation of aromatic molecules. 

Everything that has been presented above establishes an exciting ground to explore aromatic chemistry in cold environments. Furthermore, the detection of \ce{C6H5CN}, \citet{McGuire2018}, also presents an intensive, but unsuccessful, search for other aromatic molecules, including, but not limited to, furan, pyrrole, and pyridine. The detection of cyanocyclopentadiene \citep{McCarthy2020} also raises the question of why aromatic heterocycles, such as pyridine or pyrrole, are seemingly absent from observations. Both \citet{McGuire2018} and \citet{McCarthy2020} hypothesize that chemically active radicals such as \ce{NH} and \ce{NH2} might be in lower abundance than the inert \ce{N2} or \ce{NH3}. \ce{NH} and \ce{NH2} reaction with butadiene is postulated as a possible route for the formation of pyrrole and similarly, pyridine \citep{McCarthy2021}. An alternative explanation points to the CN radical reacting with aromatic material as a possible chemical conversion route \citep{McCarthy2020, Cooke2020}. The latter argument on chemical conversion must also hold for other reactive species presenting barrierless pathways or chemical reduction with hydrogen (H) atoms. H atoms possess the unique trait of being able to tunnel effectively through potential energy barriers, increasing the probability of interstellar chemical reactions. Hydrogenations via quantum tunneling have been extensively studied in the literature, both experimentally and theoretically \citep{Goumans2010, Meisner2017, Lamberts2017, Oba2018, Alvarez-Barcia2018, Nguyen2020, Molpeceres2021} to mention a few. The high abundance of atomic hydrogen in astronomical environments encouraged us to study the possible outcomes of the interaction of H atoms with aromatic molecules. Hydrogenation of benzene mediated by tunneling has been previously postulated as an effective process under astronomical conditions \citep{Goumans2010}.  

With this paper, we have two intentions: Firstly, we want to evaluate the viability of the hydrogenation of simple aromatic archetypes containing heteroatoms in their aromatic skeleton. Secondly and related, we want to evaluate the influence of the heteroatom on the reactivity of the archetypes. Both questions are addressed from a computational standpoint. The answers to these questions will help astronomers identify possible targets in present and future surveys looking for aromatic molecules in the interstellar medium. As archetypes for our study, we have selected all aromatic six-member and five-member rings containing heteroatoms with significant abundance in astrochemical models \citep{Asplund2006}. These include pyridine, pyrrole, furan, thiophene, silabenzene, and phosporine. The list of molecules can be found in Figure \ref{fig:structures}. Previous studies about the synthesis of heterocycles under astrophysical conditions showed that furan, tiophene and pyrrol must prevail under astronomical conditions \citep{Lattelais2010}. Gas phase synthetic routes for pyridine compatible with astrophysical conditions were also investigated in the past \citep{Anders1974,Parker2015, Parker2017}, as well as the stability of pyridine derivatives and other N heterocycles \citep{Johansen2021, Rap2020, Hendrix2020}. Synthesis of phosphorine under astrophysical conditions has also received recent attention in the literature, due to a bottom-up barrieless pathway to form it \citep{Fioroni2019} and recently, spectroscopic properties of polycyclic aromatic phosphorus heterocycles (PAPHs) were simulated \citep{Oliveira2020}. Silicon bearing heterocycles other than silabenzene have also been studied in the literature \citep{Parker2013, Fortenberry2018, Barrales2019},  although we have not found any reference to the synthesis of silabenzene under astrophysical conditions in the astrochemical literature. The archetypes' abundances in astrophysical environments will differ by several orders of magnitude due to the initial atomic abundances in dense clouds. 

In this paper, we report reaction energies, activation energies, and bimolecular reaction rate constants for each archetype reacting with H atoms. Both \ce{H} additions and \ce{H2} abstraction processes were investigated. Additionally, for fast enough reactions, we have studied the outcomes of a second hydrogenation process. The paper is divided as follows: In the first section, we present the theoretical protocol we have used to study the hydrogenation of our selected archetypes. After that, we revisit our prior results on the hydrogenation of benzene \citep{Goumans2010} and extend them, serving as a benchmark and starting point for the hydrogenation of heterocycles. Then we present the results for each of the considered archetypes, serving the last section to contextualize our results in an astronomical picture and gather the main conclusions of the present work.

\section{Methods}

\begin{figure}
\centering
\includegraphics[width=0.4\textwidth]{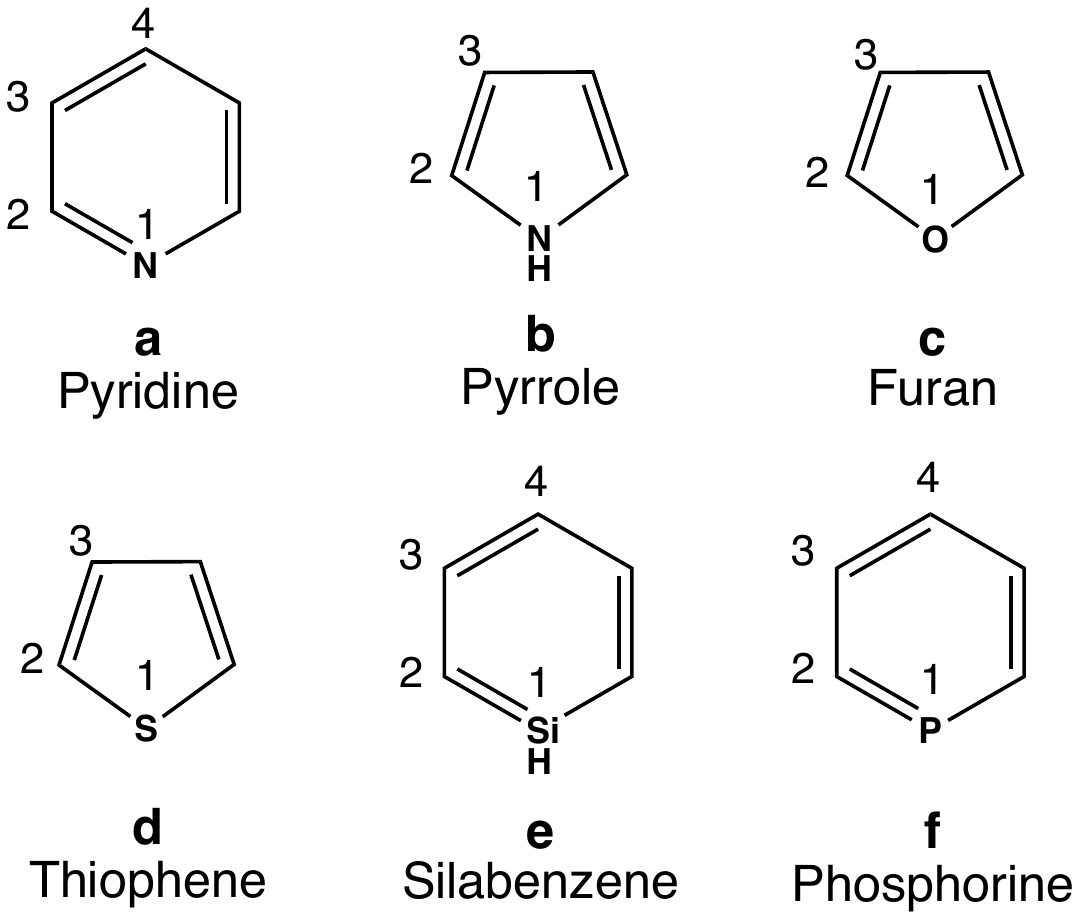}
\caption{List of structures considered in this work and associated positions where hydrogenation channels were evaluated.}
\label{fig:structures}
\end{figure}

Hydrogenation channels were modeled employing density functional theory (DFT) for all structures. According to a previous benchmark used for the hydrogenation of benzene \citep{Goumans2010}, the MPWB1K \citep{Zhao2004} exchange and correlation functional was selected. This study found that MPWB1K yielded comparable results to the more expensive CCSD(T)/CBS level. We have kept this selection, but in this study, we have increased the basis set size, employing the def2-TZVP \citep{Weigend2005} in our work. The combination of exchange and correlation functional and basis set is abbreviated as MPWB1K/def2-TZVP. Electronic structure calculations were run using the Gaussian16 code \citep{g16}. Hydrogenation reactions were sampled in each possible position of every molecule under consideration. Additionally, we repeated the study for benzene, now including the bigger basis set. The protocol we have followed to characterize all the possible chemical reactions is the same for all archetypes. First, from each relaxed structure of the archetypes, we have performed exploratory potential energy surface (PES) scans, restraining the reaction coordinate of interest. In the case of 6-member ring molecules (i.e. \textbf{a}, \textbf{e}, and \textbf{f}) depicted in Figure \ref{fig:structures}, we analyze hydrogenations in four different positions, namely the heteroatom (\textbf{1}), the ortho-position (\textbf{2}), the meta-position (\textbf{3}), and the para-position (\textbf{4}). In the case of the archetypes containing 5-membered rings (\textbf{b}, \textbf{c}, and \textbf{d}) the number of positions is reduced to three. 
Both hydrogen additions and \ce{H2} abstractions were sampled. We can discern between exothermic and endothermic processes as well as processes with and without a barrier from this initial exploration. Endothermic processes are discarded based on the low-temperature conditions of astronomical environments. All the \ce{H2} abstractions were found to be endothermic with only two exceptions for silabenzene and pyrrole.

We optimized the transition state for reactions presenting an activation barrier using the dimer method  \citep{Henkelman1999, Kastner2008}. Activation energies ($U_\text{a}$, corrected for zero-point energy) are calculated as the energy difference between reactant and transition state. Rate constants for activated processes were calculated using transition state theory and reduced instanton theory \citep{McConnell2017} above the crossover temperature and instanton theory \citep{lan67, mil75, col77, kae09a, Rommel2011, Rommel2011-2} below the crossover temperature in order to take quantum tunneling into account. Crossover temperatures, denoting temperatures where quantum tunneling starts to dominate, are defined as:

\begin{equation}
    T_\text{c} = \frac{\hbar \omega_{i}  }{2\pi k_\text{B}} ,
\end{equation}

where $\omega_{i}$ corresponds to the absolute value of the frequency of the transition mode. We constructed and optimized a discretized Feynman path consisting of 80 images, starting at T $\sim$ 0.7~$T_\text{c}$. A sequential cooling scheme was applied until we reached a minimum temperature of 50~K, where sufficiently good convergence of the rate constants was obtained. We doubled the number of images at 50~K to ensure convergence with the number of images in the path. Symmetry factors ($\sigma$), accounting for the degeneracy in the reaction channels were included \citep{Fernandez-Ramos2007}. 

In addition, for reactions with non-negligible rate constants in the relevant temperature range [150--50~K], or in the case of barrierless additions, we have evaluated the possibility for a second hydrogen addition (or abstraction), in order to elucidate possible candidates for interstellar detection. Thereby we define rate constants as non-negligible if they are larger than  the threshold of $k=3\times$ 10$^{-17}$ cm$^{3}$ s$^{-1}$) suggested by \citet{Goumans2010} and references therein, in most of the temperature range given above.

\begin{figure}
\centering
\includegraphics[width=0.4\textwidth]{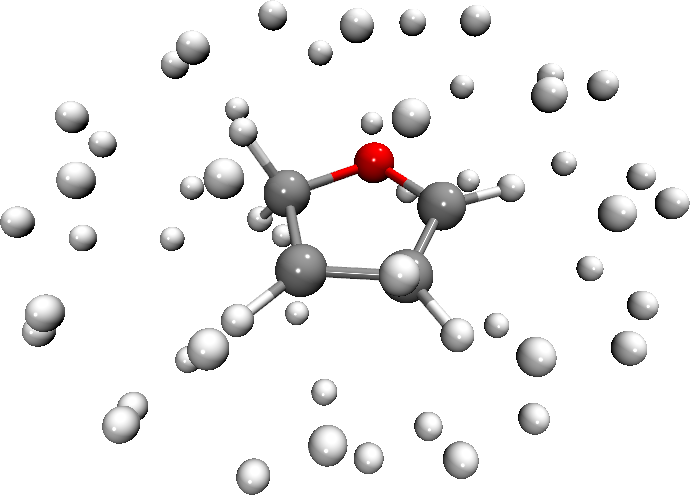}
\caption{Example of the H atoms surrounding ellipsoid, employed in the computation of branching ratios in radical-radical reactions. The figure depicts 2-hydrofuran, later investigated in this work.}
\label{fig:ellipsoid}
\end{figure}

For the study of second hydrogen additions, geometry optimizations of the hydrogenated radical coming from the first hydrogenation and an additional H atom were carried out for different starting positions (100 in each case) of the H atom spanning a distorted Fibonacci lattice, as explained in \citet{Molpeceres2021} with a tolerance value of 3~\AA~(See Fig \ref{fig:ellipsoid} for a graphic representation of this procedure). The spin state of these calculations requires to be precisely a biradical open-shell singlet at the beginning of the simulation. To achieve this, alpha and beta orbitals are mixed at the beginning of the calculation, to break the spatial spin symmetry (keyword \texttt{guess=mix} + stability analysis of the initial wavefunction in Gaussian16) \citep{EnriqueRomero2020}. Since the precise estimation of reaction parameters is not needed in this case, but rather the reaction outcome, we have reduced the basis set for these calculations to  def2-SVP \citep{Weigend2005} (MPWB1K/def2-SVP). The resulting optimized structures were inspected in each case, determining the possible outcomes of the reaction.

\section{Results}

\subsection{Hydrogenation of Benzene: Reference Calculations}
The main focus of this article is the study of hydrogenation reactions of small aromatic heterocyclic molecules.
Nevertheless, the study of the hydrogenation of benzene, a small homocyclic aromatic molecule, shall serve as a point of reference for the other reactions presented in this article because the reaction rate constants for the hydrogenation of benzene were reported previously by \citet{Goumans2010}. 

\begin{figure}
\centering
\includegraphics[width=0.4\textwidth]{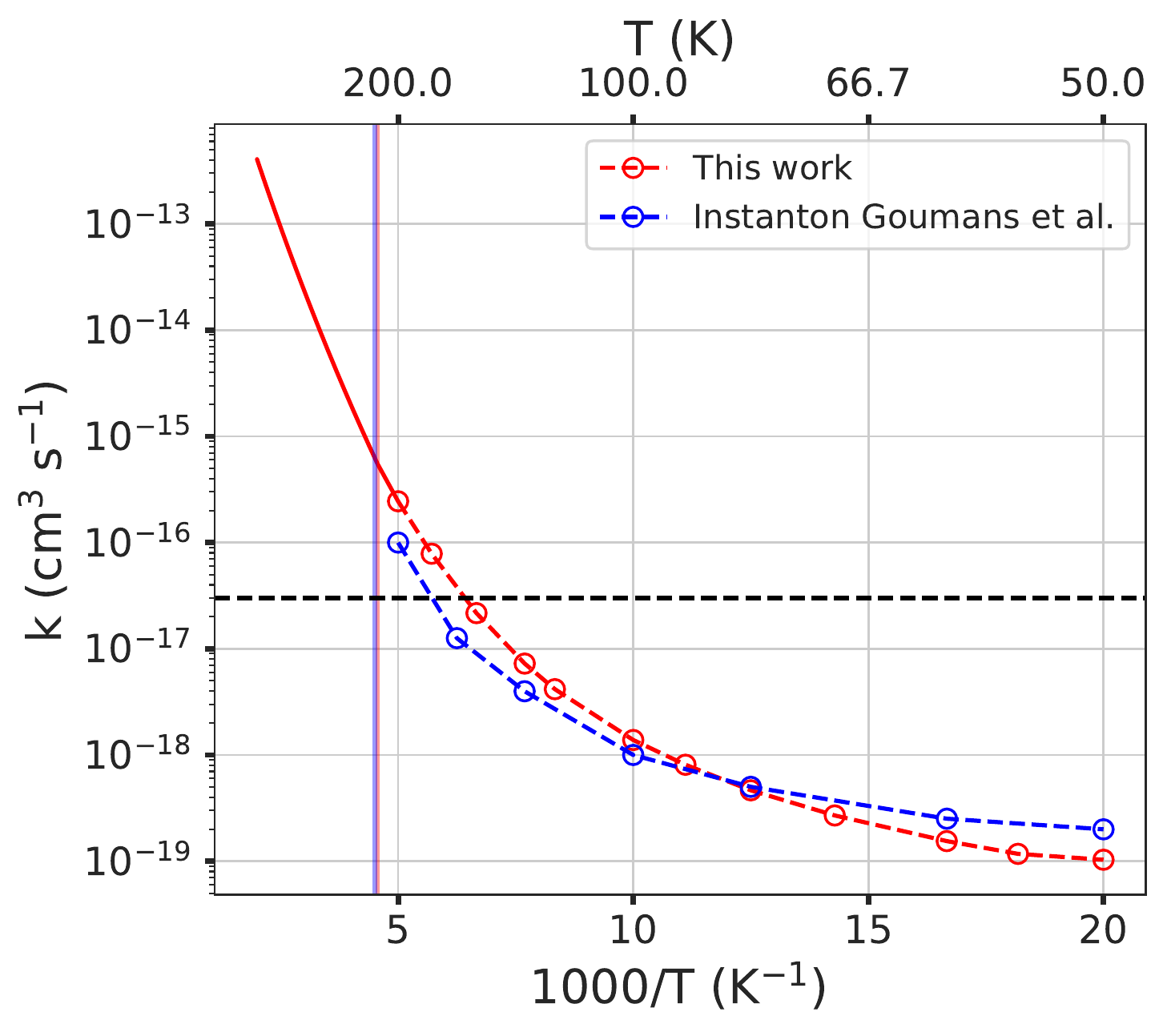}
\caption{Bimolecular reaction rate constants for the hydrogenation of benzene. The dashed horizontal line represents the tentative viability threshold presented by \citet{Goumans2010}. Vertical lines represent the crossover temperature ($T_\text{c}$) of each reaction. }
\label{fig:benzene}
\end{figure}

The hydrogenation of benzene was found to be strongly exothermic ($U_\text{r}=-89.2$ kJ mol$^{-1}$) with the activation energy being $U_\text{a}=25.8$ kJ mol$^{-1}$.
As expected, the barrier for hydrogen addition reported in \citet{Goumans2010} ($U_\text{a}=23.7$ kJ mol$^{-1}$) coincides well with the barrier found by our calculations using a larger basis.
The bimolecular rate constants for the hydrogenation of benzene are given in Fig.~\ref{fig:benzene}.
This graph shows that in general, the rate constants provided coincide rather well with the ones reported by Goumans et al. However, it can be observed that employing a bigger basis set reveals that the quantum tunnel effect affects the rate constants less pronounced than previously assumed. This is mirrored in the fact that the rate constants at low temperatures (below 70K) are slightly lower than the ones reported by \citet{Goumans2010}.

Further, a second hydrogen addition to benzene was studied. Although rate constants for the first hydrogenation of benzene are rather low, the importance of benzene as an already detected cyclic molecule \citep{Cernicharo2001, Malek2011} encourages us to evaluate a second hydrogenation process, even if the rate constants are below the tentative threshold criterion. 
Excluding the non-reactive events, 1,2-dihydrobenzene (cyclohexa-1,3-diene) and 1,4-dihydrobenzene (cyclohexa-1,4-diene) were found with a close to 1:1 ratio. 1,3-dihydrobenzene and \ce{H2} abstraction were only observed once. We observed for all archetypes that a significant portion of the geometry optimizations performed to study the second hydrogenation did not yield any hydrogen addition or abstraction but yielded the hydrogen being stuck in a pre-reactive complex.

\subsection{Hydrogenation of Pyridine}
\begin{figure}
\centering
\includegraphics[width=0.4\textwidth]{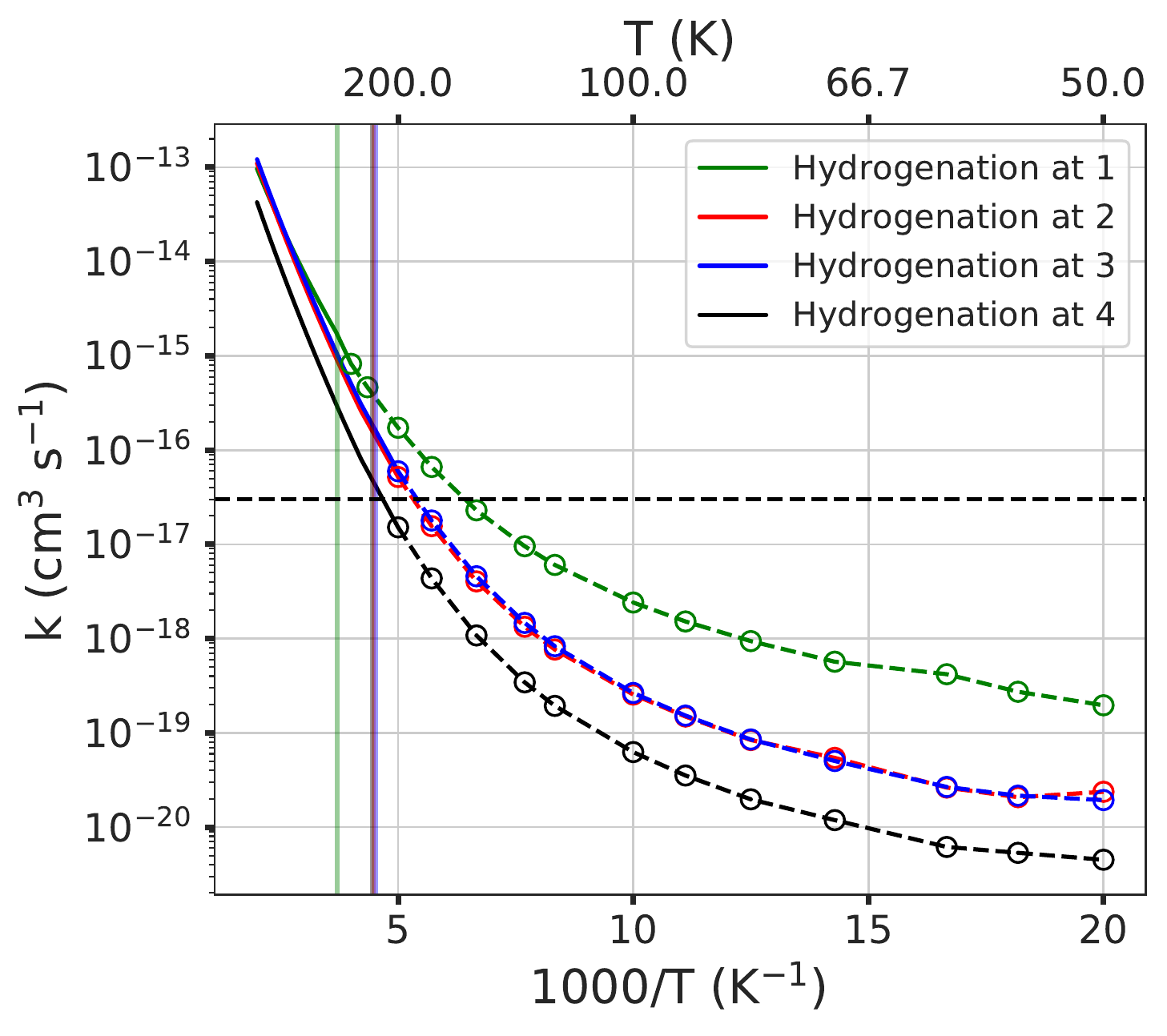}
\caption{Bimolecular reaction rate constants for the H-addition to pyridine at the 1, 2, 3 and 4 position. The dashed horizontal line represents the tentative viability threshold presented by \citet{Goumans2010}. Vertical lines represent the crossover temperature ($T_\text{c}$) of each reaction.}
\label{fig:pyridine}

\end{figure}

For pyridine (\textbf{a}) it was found that all hydrogen addition reactions are exothermic ($\textbf{1}$: $U_\text{r}=-127.4$ kJ mol$^{-1}$, $\textbf{2}$: $U_\text{r}=-88.4$ kJ mol$^{-1}$, $\textbf{3}$: $U_\text{r}= -92.0$ kJ mol$^{-1}$, $\textbf{4}$: $U_\text{r}=-88.0$ kJ mol$^{-1}$). The activation energies for the hydrogen additions indicate a preference of a hydrogenation at the nitrogen (N) atom as the activation energy for this reaction ($U_\text{a} = 25.1$ kJ mol$^{-1}$) is slightly lower than the activation energies for addition in the other positions (\textbf{2}: $U_\text{a} = 26.8$ kJ mol$^{-1}$, \textbf{3}: $U_\text{a} = 26.4$ kJ mol$^{-1}$, \textbf{4}: $U_\text{a} = 27.9$ kJ mol$^{-1}$). The similarity of the activation energies of the hydrogenation at the ortho (\textbf{2}) and meta (\textbf{3}) position suggests that the rate constants of these reactions should be nearly identical. Both, the preferential addition to the N atom as well as the similarity of the rate constants for the hydrogenations in the ortho (\textbf{2}) and meta (\textbf{3}) positions is mirrored in the reaction rate constants given in Fig \ref{fig:pyridine}.

In the temperature range 150--50~K the reaction rate constants are between $6.6 \times 10^{-17}$ and $4.5 \times 10^{-21}$ cm$^3$ s$^{-1}$ for all hydrogen addition reactions.
For the most part of this temperature range, the rate constants of all hydrogenation reactions are significantly smaller than the tentative viability threshold. Thus, no significant hydrogenation of pyridine is expected, wherefore no further investigations of this archetype were performed.

\subsection{Hydrogenation of Pyrrole}

\begin{figure}
\centering
\includegraphics[width=0.4\textwidth]{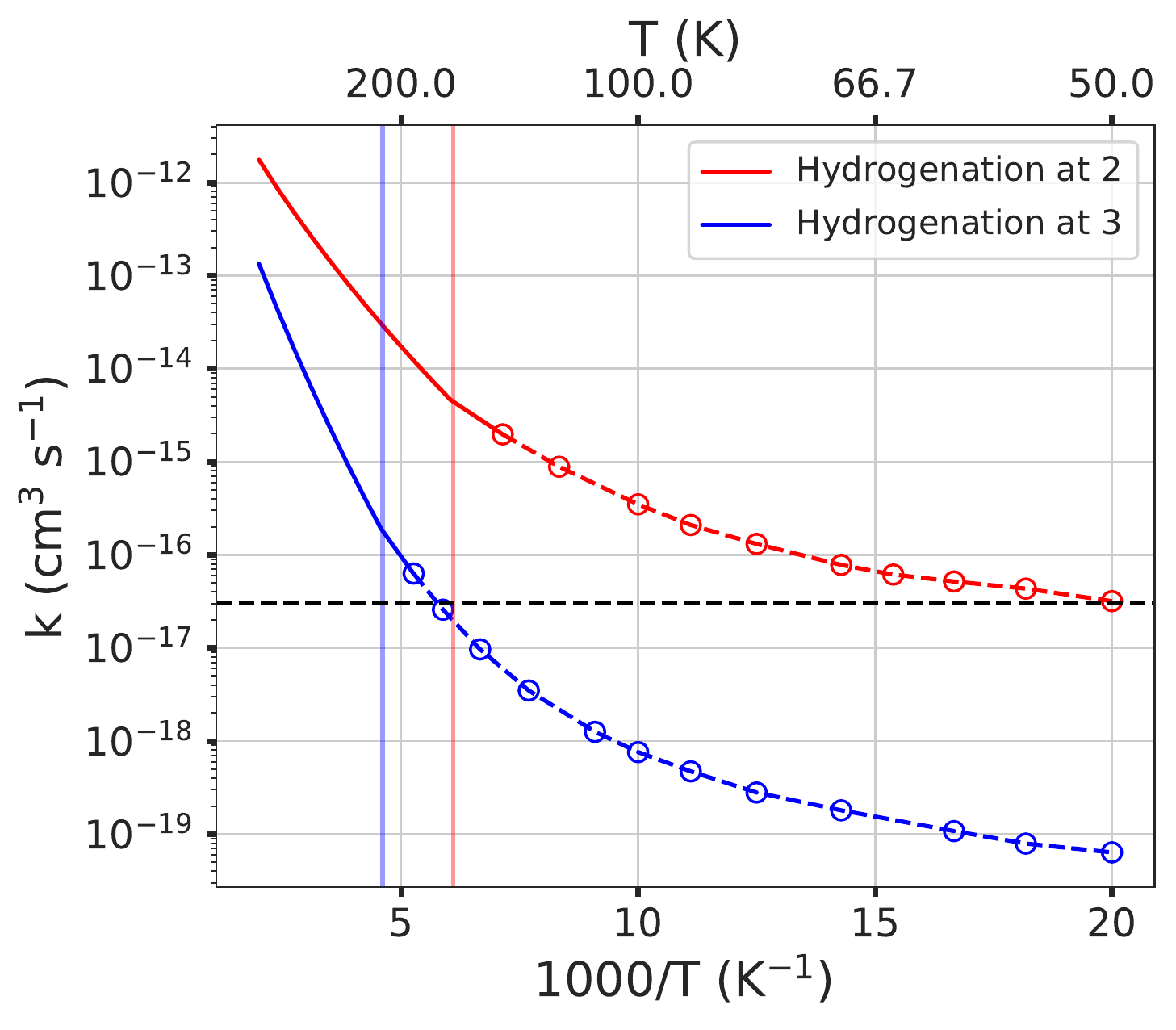}
\caption{Bimolecular reaction rate constants for the addition of H at the $2$ and $3$ positions of pyrrole. The dashed horizontal line represents the tentative viability threshold presented by \citet{Goumans2010}. Vertical lines represent the crossover temperature ($T_\text{c}$) of each reaction.}
\label{fig:pyrrole}
\end{figure}

The hydrogenation of pyrrole (\textbf{b}) occurs through H-addition reactions although \textbf{b} is one of the few sampled cases where there is a slightly exothermic \ce{H2} abstraction process (abstraction of the H attached to the N atom of pyrrole, $-17.5$ kJ mol$^{-1}$ not ZPE-corrected). Despite being exothermic, this \ce{H2} abstraction presents the highest activation barrier considered in this work of 82.0 kJ mol$^{-1}$, ZPE corrected, yielding the reaction non-competitive with the H additions. Thus, we have not deepened the discussion for the \ce{H2} abstraction channel. Both addition reactions to C are exothermic ($\textbf{2}$: $U_\text{r} = -104.6$ kJ mol$^{-1}$, $\textbf{3}$: $U_\text{r} = -67.8$ kJ mol$^{-1}$), whereas hydrogenation of the nitrogen is endothermic ($U_\text{r}$ = 82.2 kJ mol$^{-1}$). Hydrogen addition barriers point to a preferential hydrogenation at position $\textbf{2}$ ($U_\text{a}$ = 14.7 kJ mol$^{-1}$) rather than at position $\textbf{3}$ ($U_\text{a}$ = 25.5 kJ mol$^{-1}$). Conversely, rate constants vary by around 3 orders of magnitude between hydrogenation at $\textbf{2}$ and $\textbf{3}$, yielding the reaction at $\textbf{2}$ suitable for a further study of a second hydrogenation while discarding the reaction at $\textbf{3}$. The absolute value of the rate constants at 50~K is 3.2\e{-17} cm$^{3}$ s$^{-1}$ for the hydrogenation at $\textbf{2}$ yielding 2-hydropyrrole and 6.4\e{-20} cm$^{3}$ s$^{-1}$ for the one at $\textbf{3}$ leading to 3-hydropyrrole. Arrhenius plots for this reaction are presented in Fig. \ref{fig:pyrrole}.

Given these rate constants, further hydrogenation is expected in 2-hydropyrrole. The calculations for the second hydrogenation showed a variety of possible outcomes: 2,3-dihydropyrrole, 2,5-dihydropyrrole and \textbf{b} + \ce{H2}. We have not found any 2,4-dihydropyrrole in our simulations. Further, we have not found any preference towards the formation of a specific species of the three possible ones presented above. It is also important to mention that \ce{H2} abstraction takes place only at the --\ce{CH2} moiety and that we have not observed any reaction with other hydrogen atoms of the structure (i.e. the N-H group). 

\subsection{Hydrogenation of Furan}

Furan (\textbf{c}) reacts with hydrogen preferentially at the carbon atoms, with the H addition to the oxygen atom being heavily endothermic ($U_\text{r}$ = 158.8 kJ mol$^{-1}$). On the other hand, addition to the carbon atoms is exothermic ($\textbf{2}$: $U_\text{r} = -132.1$ kJ mol$^{-1}$, $\textbf{3}$: $U_\text{r} = -83.2$ kJ mol$^{-1}$) but presents activation energies. The magnitude of these activation energies is $U_\text{a}$ = 15.3 kJ mol$^{-1}$ for the hydrogenation at position $\textbf{2}$ and of $U_\text{a}$ = 27.3 kJ mol$^{-1}$ for the $\textbf{3}$ counterpart. 

\begin{figure}
\centering
\includegraphics[width=0.4\textwidth]{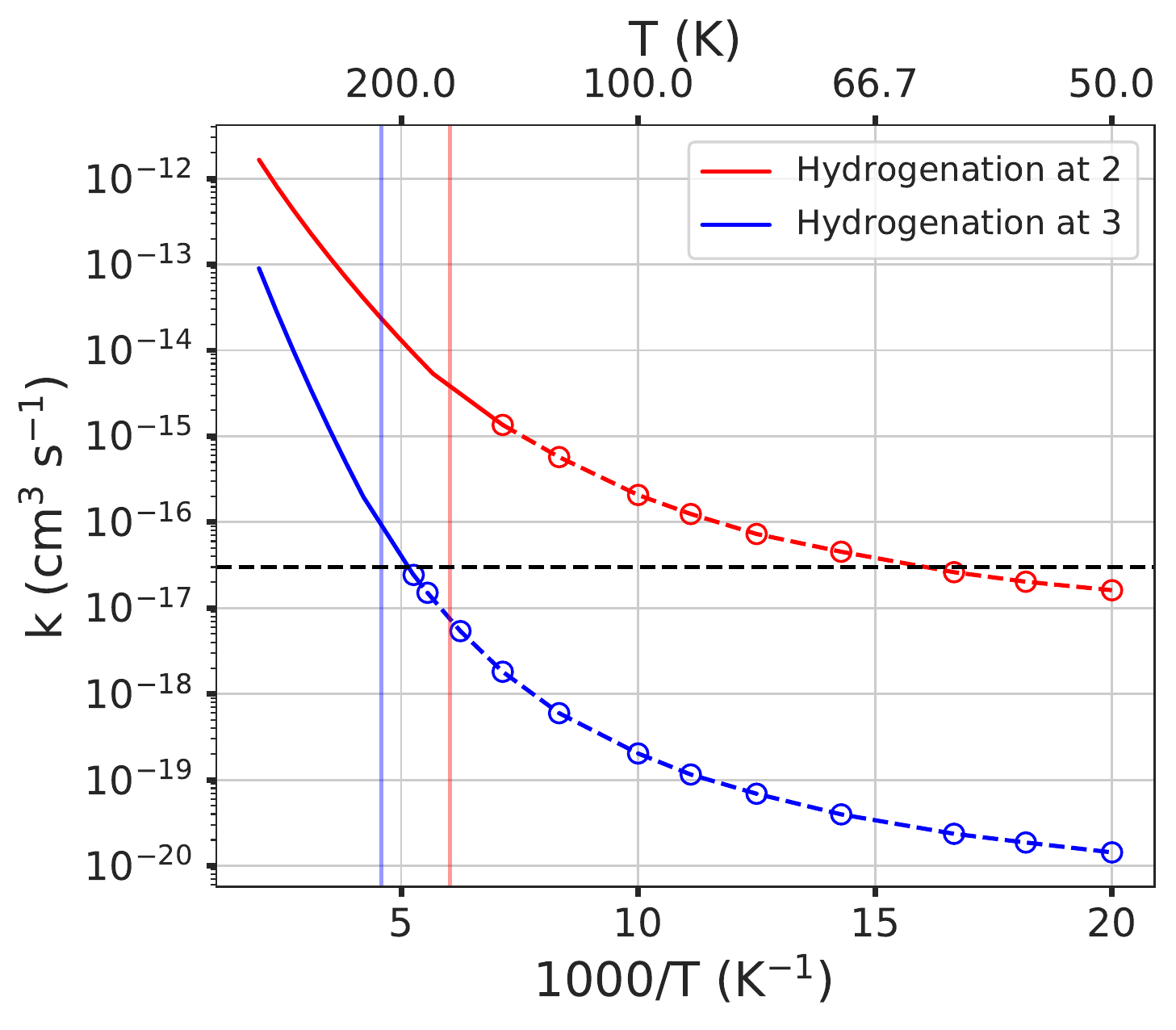}
\caption{Bimolecular reaction rate constants for the addition of H at the $2$ and $3$ position of furan. The dashed horizontal line represents the tentative viability threshold presented by \citet{Goumans2010}.}
\label{fig:furan}
\end{figure}

From the significant gap in activation energies, a similar gap in instanton rate constants is expected. This is confirmed by our calculations, showing a difference of $\sim$~3 orders of magnitude between both rate constants at 50~K. Bimolecular rate constants for the reactions are presented in Fig~\ref{fig:furan}. Hydrogenation at position $\textbf{2}$ of \textbf{c} is expected due to sizable rate constants across the whole temperature range under consideration. By contrast, no significant hydrogenation of \textbf{c} in position $\textbf{3}$ is expected between 200 and 50~K. Therefore, it is viable to assume that 2-hydrofuran is further reacting via barrierless recombination with an additional hydrogen atom. 

We have computed the second hydrogenation channels for this reaction. We obtain a similar amount of 2,5- and 2,3-dihydrofuran. As in the case of \textbf{b}, 2,4-dihydrofuran was not observed. Surprisingly, almost no \ce{H2} abstraction reactions were obtained (only 2 out of 100 trajectories). However, not finding 2,4-dihydrofuran as a possible outcome in our calculations, and finding that a significant fraction of the second addition results in 2,3- and 2,5-dihydrofuran suggests that these archetypes could be present in astrophysical environments.

\subsection{Hydrogenation of Thiophene}

\begin{figure}
\centering
\includegraphics[width=0.4\textwidth]{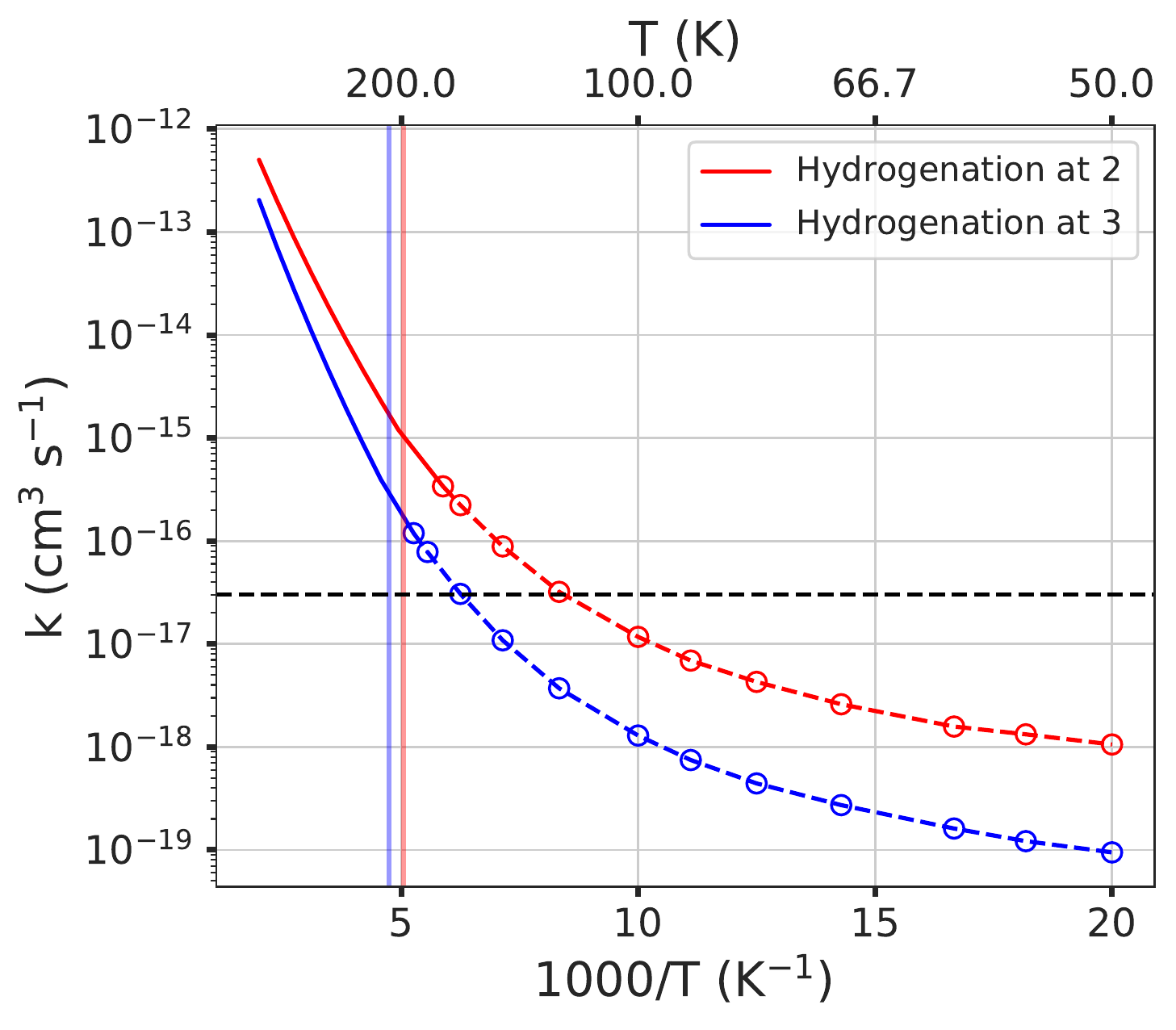}
\caption{Bimolecular reaction rate constants for the addition of H at the $2$ and $3$ position of thiophene. The dashed horizontal line represents the tentative viability threshold presented by \citet{Goumans2010}. Vertical lines represent the crossover temperature ($T_\text{c}$) of each reaction.}
\label{fig:tiophen}
\end{figure}

Thiophene (\textbf{d}) was found to be an archetype that is very resilient to hydrogenation. Under our conditions no significant chemical conversion has been found. Hydrogenation of the sulfur (S) atom is unfavorable owing its endothermicity. Hydrogenation at the carbon atoms is, on the other hand, exothermic ($\textbf{2}$: $U_\text{r} = -133.8$ kJ mol$^{-1}$, $\textbf{3}$: $U_\text{r} = -91.9$ kJ mol$^{-1}$), and comparatively more exothermic that hydrogenation of the furan (\textbf{c}) molecule. However, reaction barriers are higher than in the case of \textbf{c} for the $\textbf{2}$ position ($\textbf{2}$: $U_\text{a}$ = 20.4 kJ mol$^{-1}$) while lower in the $\textbf{3}$ position ($\textbf{3}$: $U_\text{a}$ = 24.0 kJ mol$^{-1}$). Instanton reaction rate constants are presented in Fig.~\ref{fig:tiophen}. On the basis of the instanton rate constants we can confirm that the increased activation energy for the hydrogenation at position $\textbf{2}$ precludes any significant reactivity, with rate coefficients in the order of $\sim$ 8\e{-17}--1\e{-18} cm$^{3}$s$^{-1}$ in the range between 150--50~K. Since no significant hydrogenation is expected for the first H addition, we have not deepened the description of this archetype.

\subsection{Hydrogenation of Silabenzene}

\begin{figure}
\centering
\includegraphics[width=0.4\textwidth]{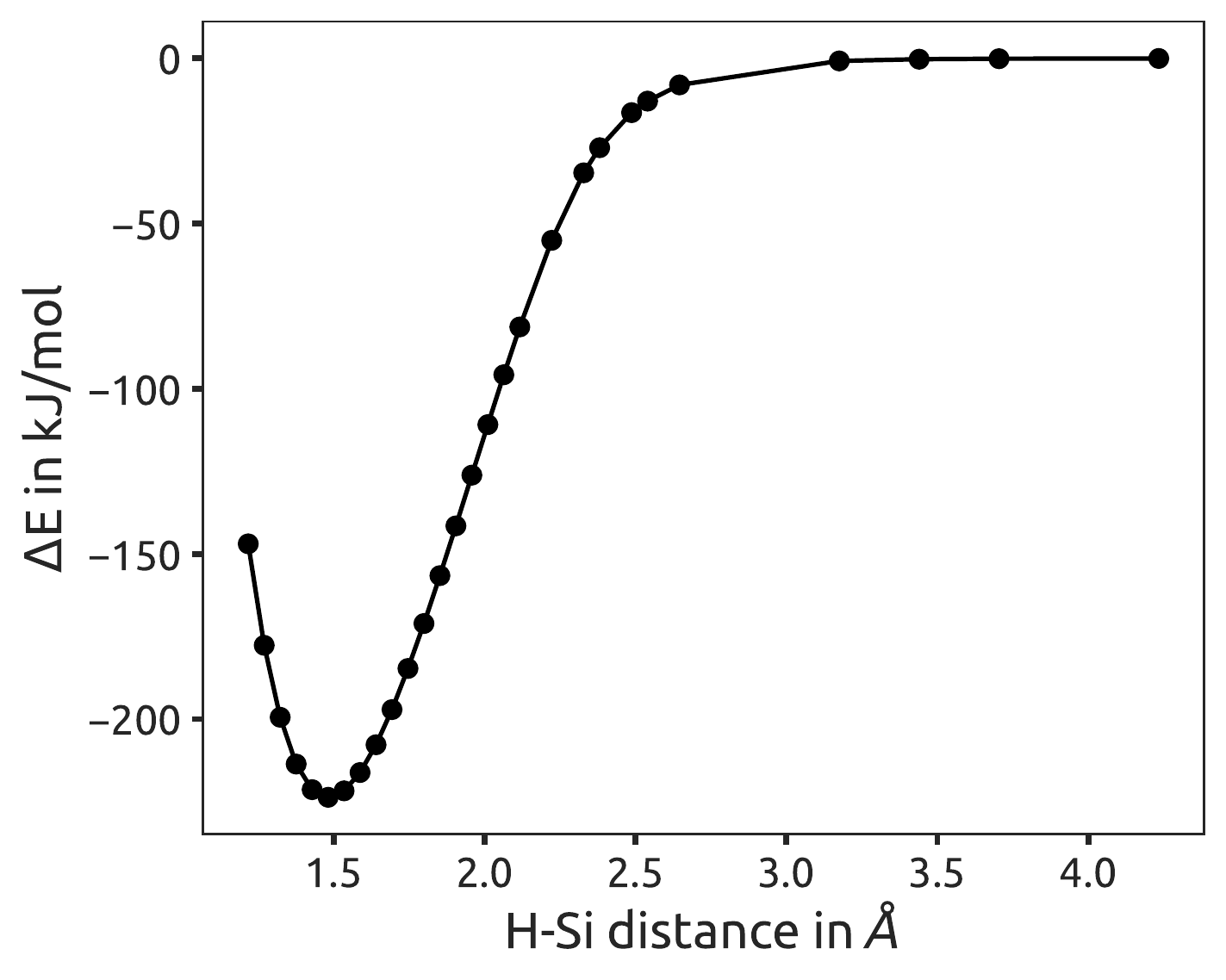}
\caption{Potential energy scan for the addition of one H atom to Silicon in silabenzene. Zeropoint energies are not included in the scan. Points represent explicit energies calculated in the scan. Lines are a guide to the eye.}
\label{fig:scan_si}
\end{figure}

As in the case of pyridine (\textbf{a}), silabenzene (\textbf{e}) can be potentially hydrogenated through four different channels. Although we sampled all positions, we were only able to produce trustworthy potential energy scans with the hydrogen added to the \textbf{1}, \textbf{3}, and \textbf{4} positions. All attempts to converge a proper 2-hydrosilabenzene molecule ended in 1-hydrosilabenzene, which in turn is formed via a barrierless path (see Fig. \ref{fig:scan_si}), indicating that \textbf{e} is a more reactive molecule in comparison with the previous archetypes. Assuming a formation of a pre-reactive complex between the H atom and \textbf{e}, other reaction pathways presenting activation barriers (\textbf{3}: $U_\text{a}$ = 25.2 kJ mol$^{-1}$, \textbf{4}: $U_\text{a}$ = 16.9 kJ mol$^{-1}$) are non-competitive against the barrierless channel. Therefore, we continued the description of the system considering a hydrogenation only at \textbf{1}. The reaction exothermicity is $U_\text{r} = -206.2$ kJ mol$^{-1}$. In addition, all \ce{H2} abstractions are endothermic except for the one at \textbf{1} with an activation barrier of $U_\text{a} = 17.1$ kJ mol$^{-1}$. Despite the fact that the barrier is of similar magnitude as the H additions presented in this paper, the presence of a barrierless path at the same atom essentially precludes it from happening.

The addition of a second hydrogen atom was sampled only considering a prior hydrogenation step of the silicon (Si) atom (position \textbf{1}). Under these conditions, we found that the second hydrogenation occurs via a barrierless path mostly in the carbon next to the Si atom (\textbf{2}, forming 1,2-dihydrosilabenzene) and in the opposite side (\textbf{4}, forming 1,4-dihydrosilabenzene), with roughly a 50 \% / 50 \% occurrence between them in the reactive cases. Rarer in our sampled cases (100 trajectories of an equispaced ellipsoid) are \ce{H2} abstraction events with just one occurrence. No 1,3-dihydrosilabenzene was obtained in our simulations.

\subsection{Hydrogenation of Phosphorine}

Phosphorine (\textbf{f}) can be, as pyridine (\textbf{a}) or silabenzene (\textbf{e}), hydrogenated in four different positions. However, it was found in our studies, that the hydrogen addition reaction to the phosphorus (P) atom (\textbf{1}) is nearly barrierless (\textbf{1}: $U_\text{a}$= 2.1 kJ mol$^{-1}$) and strongly exothermic (\textbf{1}: $U_\text{r}= -125.4$ kJ mol$^{-1}$). Assuming that a pre-reactive complex between \textbf{f} and the H atom is formed, the activation and reaction energies for the other three hydrogenation reactions are for addition at \textbf{2}: $U_{\text{a}}$=21.1 kJ mol$^{-1}$, $U_{\text{r}}=-126.3$ kJ mol$^{-1}$; at \textbf{3}: $U_{\text{a}}$=26.3 kJ mol$^{-1}$, $U_{\text{r}}=-92.1$ kJ mol$^{-1}$; and at \textbf{4}: $U_{\text{a}}$=20.0 kJ mol$^{-1}$, $U_{\text{r}}=-126.2$ kJ mol$^{-1}$.

\begin{figure}
\centering
\includegraphics[width=0.4\textwidth]{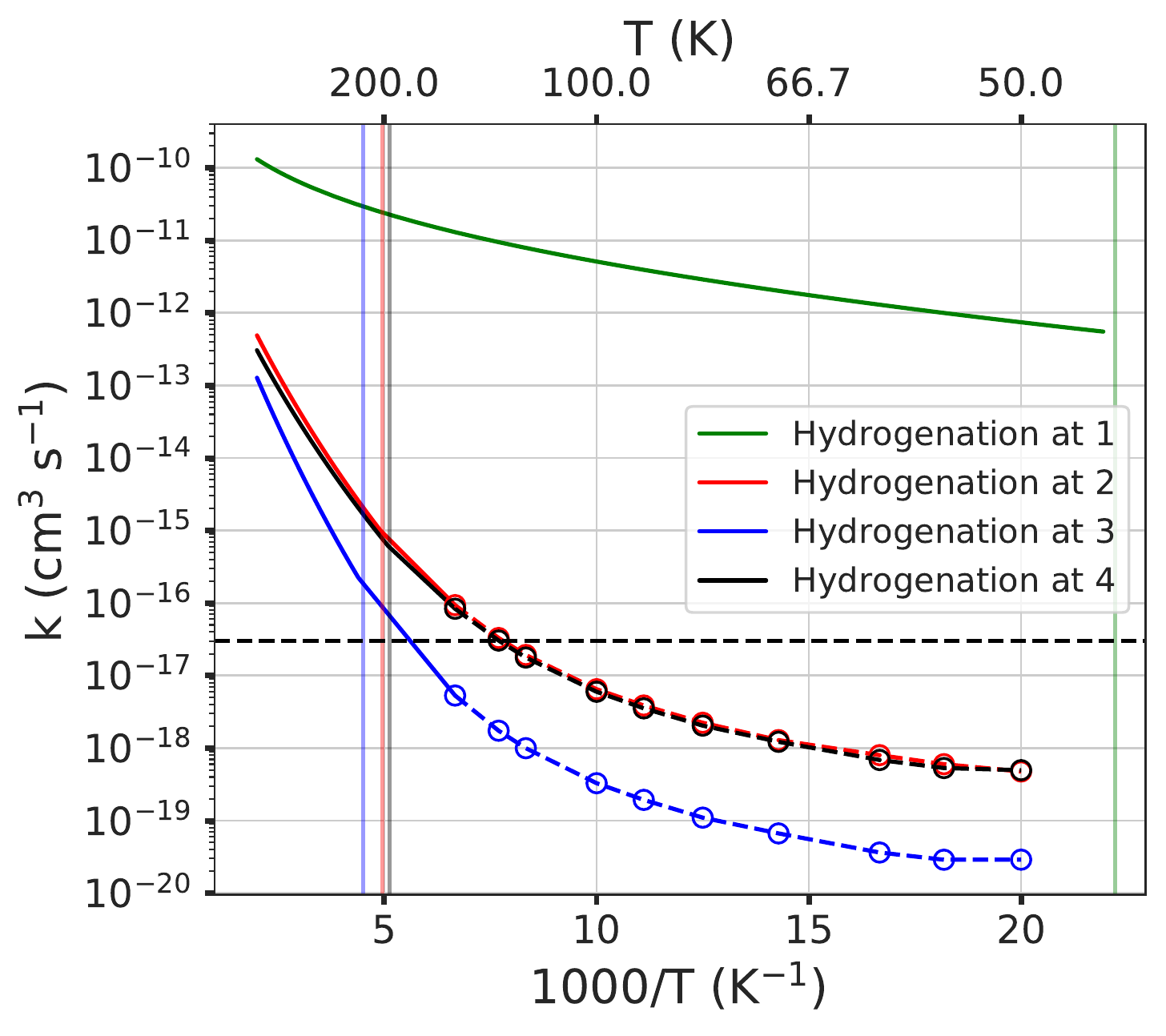}
\caption{Bimolecular reaction rate constants for the addition of H to phosphorine at the 1, 2, 3 and 4 position. The dashed horizontal line represents the tentative viability threshold presented in \citet{Goumans2010}. Vertical lines represent the crossover temperature ($T_\text{c}$) of each reaction.}
\label{fig:phosphorine}

\end{figure}

Since the activation energy for the hydrogenation of the P atom (\textbf{1}) is about ten times lower than the activation energies for the other hydrogenation reactions (\textbf{2}-\textbf{4}), this hydrogenation reaction is assumed to dominate all other hydrogenations. 
The bimolecular reaction rate constants given in Fig.~\ref{fig:phosphorine} confirm this assumption as the rate coefficients for the hydrogenation of the P atom (\textbf{1}) are over the whole temperature range 4--7 orders of magnitude larger than the rate coefficients for the other hydrogenation reactions. All \ce{H2} abstraction reactions were found to be endothermic.

Further, the addition of a second hydrogen atom to \textbf{f} was studied. Based on the rate constants it was assumed that \textbf{f} was hydrogenated in position $\textbf{1}$ during the first hydrogenation step.
It was found that the second hydrogenation occurs most often at the P atom (\textbf{1}), forming 1,1-dihydrophosphorine and at position $\textbf{4}$ yielding 1,4-dihydrophosphorine. Only one trajectory showed the formation of 1,2-dihydrophosporine and no single trajectory ended forming 1,3-dihydrophosporine. Compared to \textbf{e}, \ce{H2} abstractions were found more frequently, in about a 15 \% of the trajectories. In our simulations only \ce{H2} abstractions at the P atom forming \textbf{f} + \ce{H2} were observed. 



\section{Discussion}

\begin{table*}
    \caption{Energetics of the most-likely hydrogenations of aromatics studied in this work.}
    \label{tab:summary}
    \centering
    \resizebox{\linewidth}{!}{
    \begin{tabular}{ccccccc}
    \hline
        Molecule & Hydrogenation product & $U_{\text{a}}$ (kJ mol$^{-1}$) & $U_{\text{r}}$ (kJ mol$^{-1}$) & Second H ? &  Product of 2nd hydrogenation & \ce{H2} abstraction?\\
    \hline
        Benzene  & hydrobenzene & 25.8 & $-$89.2 & Yes & 1,2-dihydrobenzene \& 1,4-dihydrobenzene & Rare\\ 
        Pyridine & 1-hydropyridine & 25.1 & $-$127.4 & No & n/a & n/a \\
        Pyrrole & 2-hydropyrrole & 14.7 & $-$104.6 & Yes & 2,3-dihydropyrrole \& 2,5-dihydropyrrole & Yes \\
        Furan   & 2-hydrofuran & 15.3 & $-$132.1 & Yes & 2,3-dihydrofuran \& 2,5-dihydrofuran & Rare \\
        Thiophene & 2-hydrothiophene & 20.4 & $-$133.8 & No & n/a & n/a \\
        Silabenzene & 1-hydrosilabenzene & barrierless & $-$206.2 & Yes & 1,2-dihydrosilabenzene \& 1,4-dihydrosilabenzene & Rare \\
        Phosphorine & 1-hydrophosphorine & 2.1 & $-$125.4 & Yes & 1,1-dihydrophosphorine \& 1,4-dihydrophosphorine & Yes \\

    \hline
    \end{tabular}
    }
\end{table*}

Our simulations show a lack of significant trends across the whole sequence of studied heterocycles. In Table~\ref{tab:summary} we have gathered the main results of our study, grouping the outcomes of the first hydrogenation that showed the smallest activation barrier, as well as possible outcomes of the second hydrogenation processes. 

Almost all H addition reactions investigated in this study are exothermic, with the exceptions of H addition to the heteroatoms in pyrrole (\textbf{b}), furan (\textbf{c})  and thiophene (\textbf{d}). The exothermicity of the reactions roughly varies between 100--200~kJ mol$^{-1}$, which complicates the extraction of a significant trend in this quantity. Similarly, activation energies are in the range of 15--26 kJ mol$^{-1}$, when present (H addition to silabenzene (\textbf{e}) and phosphorine (\textbf{f}) are barrierless and almost barrierless, respectively). This big difference in activation barriers renders some H additions very slow (\textbf{a} and \textbf{d}) while some others are significantly faster and potentially viable on astronomical timescales (\textbf{c} and \textbf{b} most notably in addition to the barrierless processes). 

Another trend we can extract is that it seems that there is a ring size effect when it comes to the preferential reaction site of the first hydrogenation. All our considered five-membered rings present preferential hydrogenation at the position adjacent to the heteroatom (\textbf{2}) while the six-membered rings show  a preference for hydrogenation at the heteroatoms (\textbf{1}). Due to the small number of sampled structures we refrain from extracting any general rule of thumb with respect to ring size. A similar comment can be made with respect to a ``periodic effect''. While there is clear evidence of a significant reduction of the activation energy between Pyridine--Phosphorine and Benzene--Silabenzene for the first hydrogenation, we observe the opposite behavior for the Furan--Thiophene sequence.

\begin{figure}
\centering
\includegraphics[width=0.4\textwidth]{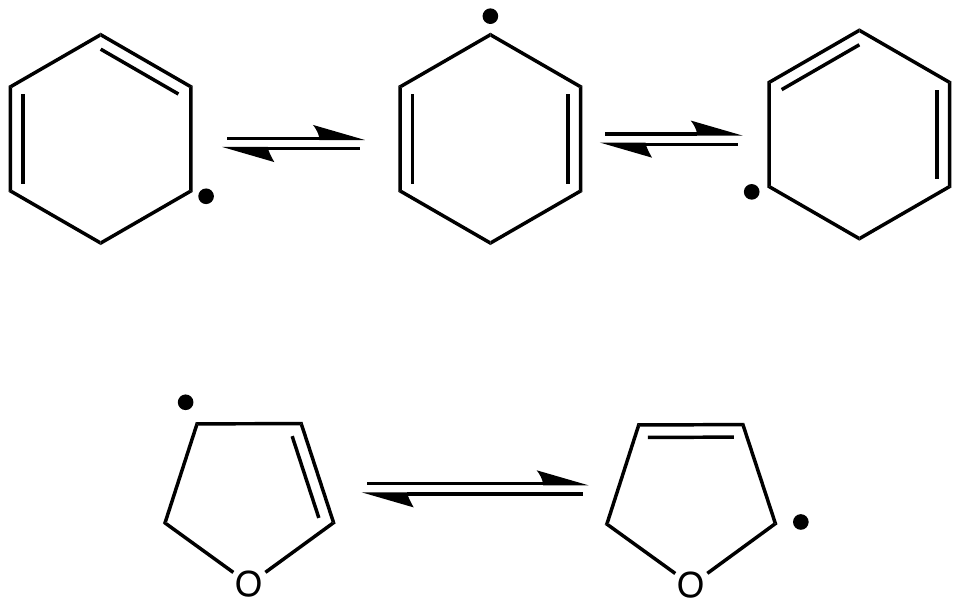}
\caption{Most important resonant forms for the  1-hydrobenzene radicals and 2-hydrofuran.}
\label{fig:resonant}
\end{figure}

The outcomes of the second hydrogenation are important because they present potential candidates for detection, given that the radicals coming from the first hydrogenation are likely too reactive to be relevant for this task. Five-membered rings show mixtures of hydrogenations in the \textbf{2},\textbf{3} and \textbf{2},\textbf{5} positions. Six-membered rings show this trend too, but in this case between \textbf{1},\textbf{2} and \textbf{1},\textbf{4}, with the exception of phosphorine (\textbf{f}) that is able to accommodate more hydrogens at the P atom owing to the hypervalence effect. It can further be observed that for all archetypes for which a second hydrogenation has been studied there is no second hydrogenation at the relative \textbf{1},\textbf{3} position (\textbf{2},\textbf{4} for five-membered rings and \textbf{1},\textbf{3} for six-membered rings). In a classical valence bond resonant picture, this is due to the impossibility of placing the single electron of the radical at that position. A graphical representation of the valence bond structures of this effect for 2-hydrofuran and 1-hydrobenzene can be found in Fig.~\ref{fig:resonant}.

The \ce{H2} abstractions after a first hydrogen addition are an important type of reaction because they can contribute to the formation of \ce{H2} in space \citep{Rauls2008, Mennella2011, Foley2018} although the magnitudes of the \ce{H2} abstraction cross sections vary by several orders of magnitude depending on the system under consideration. As an example, \citet{Mennella2011} found a significant dominance of further H additions to neutral coronene while \citet{Foley2018} found that the ratio of \ce{H} addition / \ce{H2} abstraction was closer to parity employing coronene cations. This dichotomy is mirrored in our simulations where diversity in the \ce{H2} formation appears with the different archetypes. Factors affecting \ce{H2} formation over subsequent additions require more statistical sampling and are out of the scope of this work, but could be dealt with in further work.

\section{Astrophysical Significance and Conclusions}

Our article presents a comprehensive study of the hydrogenation channels of small heterocycles. The recent wave of detections of aromatic and cyclic molecules \citep{Cernicharo2001, Malek2011, McGuire2018, McCarthy2020, Lee2021, McGuire2021} in addition to the discovery of bottom-up approaches for the synthesis of cyclic species \citep{Doddipatla2021} requires the explicit study of chemical interconversions in these molecules. The low dipole moment of many significant PAHs (i.e, benzene or naphthalene) requires the search of valid proxies for radioastronomy detection. Recently \citet{Cooke2020} showed that cyano derivatives (NC-R) are valid proxies. In fact, most of the recently detected molecules present a cyano group in their structure \citep{McGuire2018, McCarthy2020, McGuire2021}. With this work we study hydrogenations of small aromatic archetypes, postulating H as a possible radical to generate valid proxies for detection. While CN radicals are likely to present lower (or null as in the case of benzene \citet{Balucani1999, Lee2019}) activation barriers when reacting with aromatic molecules, H atoms posses a series of traits that make them interesting in this context. First, they are present in a higher abundance. Second, they are able to tunnel through potential energy barriers. Third, and most importantly, they are able to effectively diffuse on the surface of interstellar dust grains \citep{Al-Halabi2007, Hama2012, Hama2013} where all these archetypes are expected to deplete from the gas phase.

Our results show that hydrogenation is, in general, an effective process for chemical interconversion of aromatic material at low temperatures but its viability must be checked on a ``per molecule'' basis. This requirement is due to quantum tunneling efficiency being the main mechanism behind hydrogenations. Predicting the efficiency of quantum tunneling is hard at first sight since its magnitude depends on several factors, such as height and shape of the activation barrier. Quantum chemical calculations are indispensable to evaluate the rate constants of hydrogenations but should be combined with chemical modeling to ultimately predict the importance of the reactions under study.

The study of a second hydrogenation process helps us to investigate the fate of the reactive radicals generated after the first hydrogenation. The  results  presented here remain qualitative, and more trajectories should be studied to obtain statistical insight on the branching ratios of these reactions. However, from our study, we can point out several possible candidates for future astronomical surveys, namely 2,3-dihydropyrrole,  2,5-dihydropyrrole, 2,3-dihydrofuran, and 2,5-dihydrofuran. Derivatives of thiophene, silabenzene, and phosphorine were also studied in this work but the lower atomic abundances of S, Si, and P \citep{Asplund2006} in dense clouds refrain us from suggesting them as possible candidates at this stage of research. As it was mentioned in the introduction, \citet{McGuire2018} showed unsuccessful searches of pyridine and pyrrole and \citet{McCarthy2020, McCarthy2021} gave as a possible explanation the lack of \ce{NH} and \ce{NH2} radicals for their formation. One of the goals of this paper was to determine if processing with H could play a role in the apparent absence of both pyrrole and pyridine. We have confirmed the hypothesis for the former but our results remain inconclusive for the latter species. Finally, second hydrogenations can also lead to \ce{H2} abstraction reactions, but once again the efficiency of this process seems to depend on the archetype and needs to be studied on a ``per molecule'' basis, a fact that was also hinted at in the literature \citep{Mennella2011, Foley2018}.

We would like to conclude by stating that further studies are needed to evaluate possible chemical conversions of aromatics compatible with dense cloud conditions. The particular conditions in these objects summed to the handicaps in the detectability of most of them require a vast amount of data to evaluate possible candidates for detection. Bringing together experiments, observations and theoretical modeling is fundamental in the context of the exciting recent detections reported in the literature. Further iterations of the present work will include the study of reactions of archetypes with more than one aromatic cycle (e.g, quinoline derivatives), a detailed study of branching ratios, and an extension to molecules containing more than one heteroatom, for example, derivatives of oxazole and imidazole of significant importance in prebiotic chemistry. \citep{JimenezSerra2020} 

\section*{Acknowledgements}

The computer time was granted by the state of Baden-W\"urttemberg through bwHPC and the German Research Foundation (DFG) through grant no. INST 40/467-1FUGG which is greatly acknowledged. G.M thanks the Alexander von Humboldt Foundation for a post-doctoral research grant. We thank the Deutsche Forschungsgemeinschaft (DFG, German Research Foundation) for supporting this work by funding EXC 2075 - 390740016 under Germany's Excellence Strategy. We acknowledge the support by the Stuttgart Center for Simulation Science (SimTech).

\section*{Data Availability}

The data produced for this article (molecular structures, input files) will be provided on reasonable request to the corresponding author.



\bibliographystyle{mnras}
\bibliography{example} 

\begin{thebibliography}{}
\makeatletter
\relax
\def\mn@urlcharsother{\let\do\@makeother \do\$\do\&\do\#\do\^\do\_\do\%\do\~}
\def\mn@doi{\begingroup\mn@urlcharsother \@ifnextchar [ {\mn@doi@}
  {\mn@doi@[]}}
\def\mn@doi@[#1]#2{\def\@tempa{#1}\ifx\@tempa\@empty \href
  {http://dx.doi.org/#2} {doi:#2}\else \href {http://dx.doi.org/#2} {#1}\fi
  \endgroup}
\def\mn@eprint#1#2{\mn@eprint@#1:#2::\@nil}
\def\mn@eprint@arXiv#1{\href {http://arxiv.org/abs/#1} {{\tt arXiv:#1}}}
\def\mn@eprint@dblp#1{\href {http://dblp.uni-trier.de/rec/bibtex/#1.xml}
  {dblp:#1}}
\def\mn@eprint@#1:#2:#3:#4\@nil{\def\@tempa {#1}\def\@tempb {#2}\def\@tempc
  {#3}\ifx \@tempc \@empty \let \@tempc \@tempb \let \@tempb \@tempa \fi \ifx
  \@tempb \@empty \def\@tempb {arXiv}\fi \@ifundefined
  {mn@eprint@\@tempb}{\@tempb:\@tempc}{\expandafter \expandafter \csname
  mn@eprint@\@tempb\endcsname \expandafter{\@tempc}}}

\bibitem[\protect\citeauthoryear{Agundez, Cabezas, Tercero, Marcelino, Gallego,
  de Vicente  \& Cernicharo}{Agundez et~al.}{2021}]{Agundez2021}
Agundez M.,  Cabezas C.,  Tercero B.,  Marcelino N.,  Gallego J.~D.,  de
  Vicente P.,   Cernicharo J.,  2021, \mn@doi [A\&A] {10.1039/b508541a}

\bibitem[\protect\citeauthoryear{Al-Halabi \& {Van Dishoeck}}{Al-Halabi \& {Van
  Dishoeck}}{2007}]{Al-Halabi2007}
Al-Halabi A.,  {Van Dishoeck} E.~F.,  2007, \mn@doi [Mon. Notices Royal Astron.
  Soc.] {10.1111/j.1365-2966.2007.12415.x}, 382, 1648

\bibitem[\protect\citeauthoryear{{\'{A}}lvarez-Barcia, Russ, K{\"{a}}stner  \&
  Lamberts}{{\'{A}}lvarez-Barcia et~al.}{2018}]{Alvarez-Barcia2018}
{\'{A}}lvarez-Barcia S.,  Russ P.,  K{\"{a}}stner J.,   Lamberts T.,  2018,
  \mn@doi [MNRAS] {10.1093/mnras/sty1478}, 479, 2007

\bibitem[\protect\citeauthoryear{{Anders}, {Hayatsu}  \& {Studier}}{{Anders}
  et~al.}{1974}]{Anders1974}
{Anders} E.,  {Hayatsu} R.,   {Studier} M.~H.,  1974, \mn@doi [\apjl]
  {10.1086/181601}, \href
  {https://ui.adsabs.harvard.edu/abs/1974ApJ...192L.101A} {192, L101}

\bibitem[\protect\citeauthoryear{{Asplund}, {Grevesse}  \& {Jacques
  Sauval}}{{Asplund} et~al.}{2006}]{Asplund2006}
{Asplund} M.,  {Grevesse} N.,   {Jacques Sauval} A.,  2006, \mn@doi [Nuc. Phys.
  A] {10.1016/j.nuclphysa.2005.06.010}, \href
  {https://ui.adsabs.harvard.edu/abs/2006NuPhA.777....1A} {777, 1}

\bibitem[\protect\citeauthoryear{{Balucani} et~al.,}{{Balucani}
  et~al.}{1999}]{Balucani1999}
{Balucani} N.,  et~al., 1999, \mn@doi [J. Chem. Phys] {10.1063/1.480070}, \href
  {https://ui.adsabs.harvard.edu/abs/1999JChPh.111.7457B} {111, 7457}

\bibitem[\protect\citeauthoryear{Barrales-Mart{\'{i}}nez \&
  Guti{\'{e}}rrez-Oliva}{Barrales-Mart{\'{i}}nez \&
  Guti{\'{e}}rrez-Oliva}{2019}]{Barrales2019}
Barrales-Mart{\'{i}}nez C.,  Guti{\'{e}}rrez-Oliva S.,  2019, \mn@doi [MNRAS]
  {10.1093/mnras/stz2352}, 490, 172

\bibitem[\protect\citeauthoryear{Burkhardt, Loomis, Shingledecker, Lee,
  Remijan, McCarthy  \& McGuire}{Burkhardt et~al.}{2021}]{Burkhardt2021}
Burkhardt A.~M.,  Loomis R.~A.,  Shingledecker C.~N.,  Lee K. L.~K.,  Remijan
  A.~J.,  McCarthy M.~C.,   McGuire B.~A.,  2021, \mn@doi [Nature Astronomy]
  {10.1038/s41550-020-01253-4}

\bibitem[\protect\citeauthoryear{Campbell, Holz, Gerlich  \& Maier}{Campbell
  et~al.}{2015}]{Campbell2015}
Campbell E.~K.,  Holz M.,  Gerlich D.,   Maier J.~P.,  2015, \mn@doi [Nature]
  {10.1038/nature14566}, 523, 322

\bibitem[\protect\citeauthoryear{Cernicharo, Heras, Tielens, Pardo, Herpin,
  Gu{\'{e}}lin  \& Waters}{Cernicharo et~al.}{2001}]{Cernicharo2001}
Cernicharo J.,  Heras A.~M.,  Tielens A. G. G.~M.,  Pardo J.~R.,  Herpin F.,
  Gu{\'{e}}lin M.,   Waters L. B. F.~M.,  2001, \mn@doi [ApJ] {10.1086/318871},
  546, L123

\bibitem[\protect\citeauthoryear{Coleman}{Coleman}{1977}]{col77}
Coleman S.,  1977, \mn@doi [Phys. Rev. D] {10.1103/PhysRevD.15.2929}, 15, 2929

\bibitem[\protect\citeauthoryear{Cooke, Gupta, Messinger  \& Sims}{Cooke
  et~al.}{2020}]{Cooke2020}
Cooke I.~R.,  Gupta D.,  Messinger J.~P.,   Sims I.~R.,  2020, \mn@doi [ApJL]
  {10.3847/2041-8213/ab7a9c}, 891, L41

\bibitem[\protect\citeauthoryear{Doddipatla et~al.,}{Doddipatla
  et~al.}{2021}]{Doddipatla2021}
Doddipatla S.,  et~al., 2021, \mn@doi [Science Advances]
  {10.1126/sciadv.abd4044}, 7

\bibitem[\protect\citeauthoryear{Enrique-Romero et~al.,}{Enrique-Romero
  et~al.}{2020}]{EnriqueRomero2020}
Enrique-Romero J.,  et~al., 2020, \mn@doi [MNRAS] {10.1093/mnras/staa484}, 493,
  2523

\bibitem[\protect\citeauthoryear{Eriksen et~al.,}{Eriksen
  et~al.}{2020}]{Eriksen2020}
Eriksen J.~J.,  et~al., 2020, \mn@doi [J. Phys. Chem. Lett.]
  {10.1021/acs.jpclett.0c02621}, 11, 8922

\bibitem[\protect\citeauthoryear{Fern{\'{a}}ndez-Ramos, Ellingson,
  Meana-Pa{\~{n}}eda, Marques  \& Truhlar}{Fern{\'{a}}ndez-Ramos
  et~al.}{2007}]{Fernandez-Ramos2007}
Fern{\'{a}}ndez-Ramos A.,  Ellingson B.~A.,  Meana-Pa{\~{n}}eda R.,  Marques J.
  M.~C.,   Truhlar D.~G.,  2007, \mn@doi [Theor. Chem. Account.]
  {10.1007/s00214-007-0328-0}, 118, 813

\bibitem[\protect\citeauthoryear{Fioroni, Savage  \& Deyonker}{Fioroni
  et~al.}{2019}]{Fioroni2019}
Fioroni M.,  Savage R.~E.,   Deyonker N.~J.,  2019, \mn@doi [Phys. Chem. Chem.
  Phys.] {10.1039/c9cp00547a}, 21, 8015

\bibitem[\protect\citeauthoryear{Foley, Cazaux, Egorov, Boschman, Hoekstra  \&
  Schlath{\"{o}}lter}{Foley et~al.}{2018}]{Foley2018}
Foley N.,  Cazaux S.,  Egorov D.,  Boschman L.~M.,  Hoekstra R.,
  Schlath{\"{o}}lter T.,  2018, \mn@doi [Mon. Not. R. Astron. Soc.]
  {10.1093/mnras/sty1528}, 479, 649

\bibitem[\protect\citeauthoryear{Fortenberry, Novak  \& Lee}{Fortenberry
  et~al.}{2018}]{Fortenberry2018}
Fortenberry R.~C.,  Novak C.~M.,   Lee T.~J.,  2018, \mn@doi [J. Chem. Phys.]
  {10.1063/1.5043166}, 149, 024303

\bibitem[\protect\citeauthoryear{Frisch et~al.,}{Frisch et~al.}{2016}]{g16}
Frisch M.~J.,  et~al., 2016, Gaussian 16 {R}evision {C}.01

\bibitem[\protect\citeauthoryear{Goumans \& K{\"{a}}stner}{Goumans \&
  K{\"{a}}stner}{2010}]{Goumans2010}
Goumans T.~P.,  K{\"{a}}stner J.,  2010, \mn@doi [Angew. Chem. Int. Ed.]
  {10.1002/anie.201001311}, 49, 7350

\bibitem[\protect\citeauthoryear{Hama \& Watanabe}{Hama \&
  Watanabe}{2013}]{Hama2013}
Hama T.,  Watanabe N.,  2013, Chem. Rev., 113, 8783

\bibitem[\protect\citeauthoryear{Hama, Kuwahata, Watanabe, Kouchi, Kimura,
  Chigai  \& Pirronello}{Hama et~al.}{2012}]{Hama2012}
Hama T.,  Kuwahata K.,  Watanabe N.,  Kouchi A.,  Kimura Y.,  Chigai T.,
  Pirronello V.,  2012, \mn@doi [Astrophys. J.] {10.1088/0004-637X/757/2/185},
  757, 185

\bibitem[\protect\citeauthoryear{Hendrix, Bera, Lee  \& Head-Gordon}{Hendrix
  et~al.}{2020}]{Hendrix2020}
Hendrix J.,  Bera P.~P.,  Lee T.~J.,   Head-Gordon M.,  2020, \mn@doi [J. Phys.
  Chem. A] {10.1021/acs.jpca.9b11305}, 124, 2001

\bibitem[\protect\citeauthoryear{Henkelman \& Jónsson}{Henkelman \&
  Jónsson}{1999}]{Henkelman1999}
Henkelman G.,  Jónsson H.,  1999, \mn@doi [J. Chem. Phys.] {10.1063/1.480097},
  111, 7010

\bibitem[\protect\citeauthoryear{Jiménez-Serra et~al.,}{Jiménez-Serra
  et~al.}{2020}]{JimenezSerra2020}
Jiménez-Serra I.,  et~al., 2020, \mn@doi [Astrobiology]
  {10.1089/ast.2019.2125}, 20, 1048

\bibitem[\protect\citeauthoryear{Johansen, Xu, Westerfield, Wannenmacher  \&
  Crabtree}{Johansen et~al.}{2021}]{Johansen2021}
Johansen S.~L.,  Xu Z.,  Westerfield J.~H.,  Wannenmacher A.~C.,   Crabtree
  K.~N.,  2021, \mn@doi [J. Phys. Chem. A] {10.1021/acs.jpca.0c09833}, 125,
  1257

\bibitem[\protect\citeauthoryear{K\"astner \& Sherwood}{K\"astner \&
  Sherwood}{2008}]{Kastner2008}
K\"astner J.,  Sherwood P.,  2008, \mn@doi [J. Chem. Phys.]
  {10.1063/1.2815812}, 128, 014106

\bibitem[\protect\citeauthoryear{K\"astner, Carr, Keal, Thiel, Wander  \&
  Sherwood}{K\"astner et~al.}{2009}]{kae09a}
K\"astner J.,  Carr J.~M.,  Keal T.~W.,  Thiel W.,  Wander A.,   Sherwood P.,
  2009, \mn@doi [J. Phys. Chem. A] {10.1021/jp9028968}, 113, 11856

\bibitem[\protect\citeauthoryear{Lamberts \& K{\"{a}}stner}{Lamberts \&
  K{\"{a}}stner}{2017}]{Lamberts2017}
Lamberts T.,  K{\"{a}}stner J.,  2017, \mn@doi [J. Phys. Chem. A]
  {10.1021/acs.jpca.7b10296}, 121, 9736

\bibitem[\protect\citeauthoryear{Langer}{Langer}{1967}]{lan67}
Langer J.~S.,  1967, \mn@doi [Ann. Phys. (N.Y.)]
  {10.1016/0003-4916(67)90200-X}, 41, 108

\bibitem[\protect\citeauthoryear{Lattelais, Ellinger, Matrane  \&
  Guillemin}{Lattelais et~al.}{2010}]{Lattelais2010}
Lattelais M.,  Ellinger Y.,  Matrane A.,   Guillemin J.-C.,  2010, \mn@doi
  [Phys. Chem. Chem. Phys.] {10.1039/B924574J}, 12, 4165

\bibitem[\protect\citeauthoryear{{Lee}, {McGuire}  \& {McCarthy}}{{Lee}
  et~al.}{2019}]{Lee2019}
{Lee} K. L.~K.,  {McGuire} B.~A.,   {McCarthy} M.~C.,  2019, \mn@doi [Phys.
  Chem. Chem. Phys.] {10.1039/C8CP06070C}, \href
  {https://ui.adsabs.harvard.edu/abs/2019PCCP...21.2946L} {21, 2946}

\bibitem[\protect\citeauthoryear{Lee et~al.,}{Lee et~al.}{2021}]{Lee2021}
Lee K. L.~K.,  et~al., 2021, Interstellar Detection of 2-Cyanocyclopentadiene,
  C$_5$H$_5$CN, a Second Five-Membered Ring Toward TMC-1 (\mn@eprint {arXiv}
  {2102.09595})

\bibitem[\protect\citeauthoryear{Liu, Kilby, Frankcombe  \& Schmidt}{Liu
  et~al.}{2020}]{Liu2020}
Liu Y.,  Kilby P.,  Frankcombe T.~J.,   Schmidt T.~W.,  2020, \mn@doi [Nature
  Communications] {10.1038/s41467-020-15039-9}, 11, 1210

\bibitem[\protect\citeauthoryear{Malek, Cami  \& Bernard-Salas}{Malek
  et~al.}{2011}]{Malek2011}
Malek S.~E.,  Cami J.,   Bernard-Salas J.,  2011, \mn@doi [ApJ]
  {10.1088/0004-637x/744/1/16}, 744, 16

\bibitem[\protect\citeauthoryear{{Marcelino, N.}, {Tercero, B.}, {Ag\'undez,
  M.}  \& {Cernicharo, J.}}{{Marcelino, N.} et~al.}{2021}]{Marcelino2021}
{Marcelino, N.} {Tercero, B.} {Ag\'undez, M.}  {Cernicharo, J.} 2021, \mn@doi
  [A\&A] {10.1051/0004-6361/202040177}, 646, L9

\bibitem[\protect\citeauthoryear{McCarthy \& McGuire}{McCarthy \&
  McGuire}{2021}]{McCarthy2021}
McCarthy M.~C.,  McGuire B.~A.,  2021, \mn@doi [J. Phys. Chem. A]
  {10.1021/acs.jpca.1c00129}

\bibitem[\protect\citeauthoryear{McCarthy et~al.,}{McCarthy
  et~al.}{2020}]{McCarthy2020}
McCarthy M.~C.,  et~al., 2020, \mn@doi [Nature Astronomy]
  {10.1038/s41550-020-01213-y}

\bibitem[\protect\citeauthoryear{McConnell \& K{\"{a}}stner}{McConnell \&
  K{\"{a}}stner}{2017}]{McConnell2017}
McConnell S.,  K{\"{a}}stner J.,  2017, \mn@doi [J. Comp. Chem.]
  {10.1002/jcc.24914}, 38, 2570

\bibitem[\protect\citeauthoryear{McGuire, Burkhardt, Kalenskii, Shingledecker,
  Remijan, Herbst  \& McCarthy}{McGuire et~al.}{2018}]{McGuire2018}
McGuire B.~A.,  Burkhardt A.~M.,  Kalenskii S.,  Shingledecker C.~N.,  Remijan
  A.~J.,  Herbst E.,   McCarthy M.~C.,  2018, \mn@doi [Science]
  {10.1126/science.aao4890}, 359, 202

\bibitem[\protect\citeauthoryear{McGuire et~al.,}{McGuire
  et~al.}{2021}]{McGuire2021}
McGuire B.~A.,  et~al., 2021, \mn@doi [Science] {10.1126/science.abb7535}, 371,
  1265

\bibitem[\protect\citeauthoryear{Meisner, Lamberts  \& K{\"{a}}stner}{Meisner
  et~al.}{2017}]{Meisner2017}
Meisner J.,  Lamberts T.,   K{\"{a}}stner J.,  2017, \mn@doi [ACS Earth. Space.
  Chem] {10.1021/acsearthspacechem.7b00052}, 1, 399

\bibitem[\protect\citeauthoryear{Mennella, Hornek{\ae}r, Thrower  \&
  Accolla}{Mennella et~al.}{2011}]{Mennella2011}
Mennella V.,  Hornek{\ae}r L.,  Thrower J.,   Accolla M.,  2011, \mn@doi
  [Astrophys. J.] {10.1088/2041-8205/745/1/l2}, 745, L2

\bibitem[\protect\citeauthoryear{Merino et~al.,}{Merino
  et~al.}{2014}]{Merino2014}
Merino P.,  et~al., 2014, \mn@doi [Nature Communications] {10.1038/ncomms4054},
  5, 3054

\bibitem[\protect\citeauthoryear{Miller}{Miller}{1975}]{mil75}
Miller W.~H.,  1975, \mn@doi [J. Chem. Phys.] {10.1063/1.430676}, 62, 1899

\bibitem[\protect\citeauthoryear{Molpeceres \& Kästner}{Molpeceres \&
  Kästner}{2021}]{Molpeceres2021}
Molpeceres G.,  Kästner J.,  2021, \mn@doi [ApJ] {10.3847/1538-4357/abe38c},
  910, 55

\bibitem[\protect\citeauthoryear{Nguyen, Oba, Shimonishi, Kouchi  \&
  Watanabe}{Nguyen et~al.}{2020}]{Nguyen2020}
Nguyen T.,  Oba Y.,  Shimonishi T.,  Kouchi A.,   Watanabe N.,  2020, \mn@doi
  [ApJL] {10.3847/2041-8213/aba695}, 898, L52

\bibitem[\protect\citeauthoryear{Oba, Tomaru, Lamberts, Kouchi  \&
  Watanabe}{Oba et~al.}{2018}]{Oba2018}
Oba Y.,  Tomaru T.,  Lamberts T.,  Kouchi A.,   Watanabe N.,  2018, \mn@doi
  [Nat. Astron.] {10.1038/s41550-018-0380-9}, 2, 228

\bibitem[\protect\citeauthoryear{Oliveira, Molpeceres, Fantuzzi,
  Quiti{\'{a}}n-Lara, Boechat-Roberty  \& K{\"{a}}stner}{Oliveira
  et~al.}{2020}]{Oliveira2020}
Oliveira R.~R.,  Molpeceres G.,  Fantuzzi F.,  Quiti{\'{a}}n-Lara H.~M.,
  Boechat-Roberty H.~M.,   K{\"{a}}stner J.,  2020, \mn@doi [Monthly Notices of
  the Royal Astronomical Society] {10.1093/mnras/staa3460}, 500, 2564

\bibitem[\protect\citeauthoryear{Parker \& Kaiser}{Parker \&
  Kaiser}{2017}]{Parker2017}
Parker D. S.~N.,  Kaiser R.~I.,  2017, \mn@doi [Chem. Soc. Rev.]
  {10.1039/C6CS00714G}, 46, 452

\bibitem[\protect\citeauthoryear{Parker, Wilson, Kaiser, Mayhall, Head-Gordon
  \& Tielens}{Parker et~al.}{2013}]{Parker2013}
Parker D. S.~N.,  Wilson A.~V.,  Kaiser R.~I.,  Mayhall N.~J.,  Head-Gordon M.,
    Tielens A. G. G.~M.,  2013, \mn@doi [ApJ] {10.1088/0004-637x/770/1/33},
  770, 33

\bibitem[\protect\citeauthoryear{Parker, Kaiser, Kostko, Troy, Ahmed, Sun, Chen
   \& Chang}{Parker et~al.}{2015}]{Parker2015}
Parker D. S.~N.,  Kaiser R.~I.,  Kostko O.,  Troy T.~P.,  Ahmed M.,  Sun B.-J.,
   Chen S.-H.,   Chang A. H.~H.,  2015, \mn@doi [Phys. Chem. Chem. Phys.]
  {10.1039/C5CP02960K}, 17, 32000

\bibitem[\protect\citeauthoryear{Rap, Marimuthu, Redlich  \& Brünken}{Rap
  et~al.}{2020}]{Rap2020}
Rap D.~B.,  Marimuthu A.~N.,  Redlich B.,   Brünken S.,  2020, \mn@doi [J.
  Mol. Spec.] {https://doi.org/10.1016/j.jms.2020.111357}, 373, 111357

\bibitem[\protect\citeauthoryear{Rauls \& Hornek{\ae}r}{Rauls \&
  Hornek{\ae}r}{2008}]{Rauls2008}
Rauls E.,  Hornek{\ae}r L.,  2008, \mn@doi [Astrophys. J.] {10.1086/587614},
  679, 531

\bibitem[\protect\citeauthoryear{Rommel \& K{\"{a}}stner}{Rommel \&
  K{\"{a}}stner}{2011}]{Rommel2011}
Rommel J.~B.,  K{\"{a}}stner J.,  2011, \mn@doi [J. Chem. Phys.]
  {10.1063/1.3587240}, 134, 184107

\bibitem[\protect\citeauthoryear{Rommel, Goumans  \& K\"astner}{Rommel
  et~al.}{2011}]{Rommel2011-2}
Rommel J.~B.,  Goumans T.~P.,   K\"astner J.,  2011, \mn@doi [J. Chem. Theory
  Comp.] {10.1021/ct100658y}, 7, 690

\bibitem[\protect\citeauthoryear{Tielens}{Tielens}{2008}]{Tielens2008}
Tielens A.~G.,  2008, \mn@doi [Annu. Rev. Astron. Astrophys.]
  {10.1146/annurev.astro.46.060407.145211}, 46, 289

\bibitem[\protect\citeauthoryear{Weigend \& Ahlrichs}{Weigend \&
  Ahlrichs}{2005}]{Weigend2005}
Weigend F.,  Ahlrichs R.,  2005, \mn@doi [Phys. Chem. Chem. Phys.]
  {10.1039/b508541a}, 7, 3297

\bibitem[\protect\citeauthoryear{Zhao \& Truhlar}{Zhao \&
  Truhlar}{2004}]{Zhao2004}
Zhao Y.,  Truhlar D.~G.,  2004, \mn@doi [J. Phys. Chem. A] {10.1021/jp048147q},
  108, 6908

\makeatother
\end{thebibliography}




\appendix




\bsp	
\label{lastpage}
\end{document}